\documentclass[twocolumn]{aastex62}

\shorttitle{The Impact of AGN-driven Winds on Metals}
\shortauthors{Choi et al.}
\def\galaxiespprox{\mathrel{\vcenter{\offinterlineskip \hbox{$>$}
    \kern 0.3ex \hbox{$\sim$}}}}
\def\lapprox{\mathrel{\vcenter{\offinterlineskip \hbox{$<$}
    \kern 0.3ex \hbox{$\sim$}}}}
\newcommand{\beq}{\begin{equation}} 
\newcommand{\eeq}{\end{equation}}

\def\eps{\epsilon}
\def\epsw{\eps_{\rm w}}

\def\Edotw{\dot{E}_{\rm w}}

\def\Sig15{\Sigma_{1.5}}

\def\kms{\rm\thinspace km~s^{-1}}

\def\cm{{\rm\thinspace cm}}

\def\kpc{{\rm\thinspace kpc}}

\def\Msun{\hbox{$\thinspace \rm M_{\odot}$}}
\def\Msunh{\hbox{$\thinspace  \it h^{\rm -1} \rm \thinspace M_{\odot} $}}

\def\pdot{\dot{p}}

\def\rg{r_{\rm g}}
\def\rh{r_{\rm h}}

\def\Mh{M_{\rm h}}

\def\Mdotacc{\dot{M}_{\rm acc}}

\def\Mdotoutf{\dot{M}_{\rm outf}}

\def\strom{Str$\rm \ddot{o}$mgren$\thinspace$}

\def\delm12{M_{12}}
\def\sigm1{\sigma(m_{1})}

\begin{document}

\title{The Impact of Outflows driven by Active Galactic Nuclei on Metals in and around Galaxies}

\correspondingauthor{Ena Choi}
\email{enachoi@kias.re.kr}
\author[0000-0002-8131-6378]{Ena Choi}
\affil{Quantum Universe Center, 
        Korea Institute for Advanced Study, 
        Hoegiro 85, Seoul 02455, Korea; enachoi@kias.re.kr}    

\author{Ryan Brennan}
\affil{Department of Physics and Astronomy, Rutgers, 
        The State University of New Jersey, \\
        136 Frelinghuysen Road, Piscataway, NJ 08854, USA}

\author{Rachel S. Somerville}
\affil{Center for Computational Astrophysics, Flatiron Institute, 162 5th Ave., 
        New York, NY 10010, USA}
\affil{Department of Physics and Astronomy, Rutgers, 
        The State University of New Jersey, \\
        136 Frelinghuysen Road, Piscataway, NJ 08854, USA}

\author{Jeremiah P. Ostriker}
\affil{Department of Astronomy, Columbia University, 
       550 West 120th Street, 
       New York, NY 10027, USA}
\affil{Department of Astrophysical Sciences, Princeton University, 
        Princeton, NJ 08544, USA}
    
\author{Michaela Hirschmann}
\affil{DARK, Niels Bohr Institute, University of Copenhagen, Jagtvej 128, 2200 Copenhagen, Denmark}

\author{Thorsten Naab}
\affil{Max-Planck-Institut f\"ur Astrophysik,
        Karl-Schwarzschild-Strasse 1, 85741 Garching, Germany}

\begin{abstract}
Metals in the hot gaseous halos of galaxies encode the history of star formation as well as the feedback processes that eject metals from the galaxies. X-ray observations suggest that massive galaxies have extended distributions of metals in their gas halos. We present predictions for the metal properties of massive galaxies and their gaseous halos from recent high resolution zoom-in simulations that include mechanical and radiation driven feedback from Active Galactic Nuclei (AGN). In these simulations, AGN launch high-velocity outflows, mimicking observed broad absorption line winds. By comparing two sets of simulations with and without AGN feedback, we show that our prescription for AGN feedback is capable of driving winds and enriching halo gas `inside-out' by spreading centrally enriched metals to the outskirts of galaxies, into the halo and beyond. The metal (iron)  profiles of halos simulated with AGN feedback have a flatter slope than those without AGN feedback, consistent with recent X-ray observations.  The predicted gas iron abundance of group scale galaxies simulated with AGN feedback is $Z_{\rm Fe} = 0.23$ $Z_{\rm Fe,\odot}$ at $0.5 r_{500}$, which is 2.5 times higher than that in simulations without AGN feedback. In these simulations, AGN winds are also important for the metal enrichment of the intergalactic medium, as the vast majority of metals ejected from the galaxy by AGN-driven winds end up beyond the halo virial radius. 
\end{abstract}

\section{Introduction} \label{sec:intro}

Our understanding of the formation and evolution of massive galaxies has made great advances in the last two decades. The current consensus is that massive elliptical galaxies formed the majority of their stars at high redshift \citep{Thomas2011,2010ApJ...709.1018V} mainly through the accretion of gas from the intergalactic medium (IGM) \citep{2005MNRAS.363....2K,Erb2007,Putman2012}, and have since grown significantly via mergers with lower mass galaxies \citep{2007ApJ...658..710N,2008ApJ...677L...5V,2009ApJ...699L.178N,2010ApJ...725.2312O}.  Despite the continual accretion of a substantial amount of gaseous material into the halo as well as the presence of plenty of hot gas available for cooling, massive elliptical galaxies show relatively low star formation rates \citep{2003MNRAS.341...33K,Renzini2006,Kormendy2009}. However, it is still unclear which physical mechanisms prevent gas from cooling and forming stars in massive galaxies.

In cosmological hydrodynamics simulations of massive galaxy formation, feedback by energetic sources is often used to suppress star formation \citep[][and references therein]{somerville2015,2017ARA&amp;A..55...59N}.  For massive galaxies, simulations  with stellar feedback but without feedback from supermassive black holes have been shown to suffer from the well-known overcooling problem with resulting stellar masses several times larger than observed \citep[e.g.][]{Balogh2001,Su2019}. Also, there are many pieces of observational evidence that support the hypothesis of AGN feedback suppressing star formation in massive galaxies  \citep{McNamara2007,2012ARA&amp;A..50..455F}. Therefore in the past two decades many simulations have started to include the feedback from supermassive black holes (BH feedback or AGN feedback) and they have shown that AGN feedback can play a crucial role in regulating the star formation in massive galaxies and suppressing the cooling of gas \citep[e.g.][]{2005ApJ...620L..79S,2013MNRAS.433.3297D,Beckmann2017}. Many simulations successfully reproduce a wide range of the basic properties of massive quiescent galaxies, such as stellar mass, number density and colors, generally attributed to an AGN that heats up the gas \citep{vogelsberger2014,Hirschmann2014,Schaye2015,Nelson2018b,2018MNRAS.479.4056W}.

This consensus is encouraging, however, AGN feedback models are still not well constrained. Most simulations on cosmological scales do not resolve the Bondi radius which is required to self-consistently model gas accretion onto black holes \citep[cf.][]{Curtis2016a,Curtis2016,Angles-Alcazar2020a}, therefore they must adopt sub-grid models for gas accretion onto black holes as well as AGN feedback to capture the unresolved physical processes \citep{2005MNRAS.361..776S,2008ApJ...676...33D,2010MNRAS.406..822M,2013AN....334..394G,2015MNRAS.452..575S,Steinborn2015,Angles-Alcazar2017a,2017MNRAS.465.3291W,Dave2019}. These subgrid implementations of AGN feedback vary widely in the literature \citep[see also][]{Wurster2013,Meece2017}. The AGN feedback recipes are currently degenerate as substantially different feedback models in simulations can result in similar galaxy stellar properties \citep[e.g.][]{Harrison2018}.

However, we can further constrain models of AGN feedback adopted in simulations using observed properties of the gaseous components of massive galaxies. It has been shown that AGN feedback has a strong effect on galaxy gas properties \citep{McCarthy2009,Dubois2012,Truong2020}, and different AGN feedback models can result in different gas properties, especially of the hot gas \citep{Yang2012,Choi2015a}. By comparing simulations with various treatments of AGN feedback (e.g. thermal or mechanical), \citet{Choi2015a} showed that depending on {\it how} the AGN feedback energy is coupled with the surrounding gas, the resulting halo X-ray luminosity can vary by three orders of magnitude, and the mechanical feedback is more successful in reproducing the observed X-ray luminosity of the hot atmospheres.

Another powerful tool to discriminate among the AGN feedback models is metallicity. Especially, metals in the hot halos of massive galaxies encode the history of star formation as well as  feedback processes that eject metals from the galaxies. X-ray observations suggest that massive galaxies and groups exhibit negative abundance gradients, being more enriched at the center, with metal abundance decreasing with increasing radius \citep{Finoguenov1999,Buote2003,Sato2009}. However, the central metal excess extends out further than the optical emission from the central galaxy, suggesting that metal enriched gas is being distributed outside of the galaxy \citep{David2008}. Some massive galaxies have halo gas whose metallicity is of the order of the metallicity within the galaxy itself. Recent UV absorption line studies probing the circumgalactic medium (CGM) of nearby galaxies also revealed a significant amount of metals at large distances from the stellar bodies of galaxies \citep{Tumlinson2013,Werk2013}. 

Meanwhile, there has been a tension between simulations and observations regarding the amount of metals in extended halos of galaxies. Simulations tend to produce galaxies with metal excesses in their centers, and deficits in the outskirts when compared with observations \citep{2003MNRAS.339.1117V,2005MNRAS.361..983R, 2008MNRAS.391..110D,2013MNRAS.432...89F}. 

AGN feedback has often been invoked to explain the observed distribution of metal-enriched gas to large radii \citep{Rebusco2006,David2008}. Some recent simulations have found that a pure thermal AGN feedback model, that locally deposits a fraction of the AGN bolometric luminosity in thermal form, has a negligible effect on halo gas metallicity \citep{Barai2013,Taylor2015a}. However, other AGN feedback implementations, such as the stochastic thermal AGN feedback model adopted in the \textsc{eagle} simulation \citep{Barnes2017}, have been shown to more efficiently distribute metal-enriched gas to large distances.  \cite{Nelson2018a} also showed that the IllustrisTNG simulations, which adopt a two-mode thermal plus kinetic AGN feedback model, better reproduce the observed metal distribution of extended warm-hot gas compared to the original Illustris simulations.  

AGN can have a dramatic effect on gaseous halos and their metallicity distributions \citep{Gaspari2012,Ciotti2016,Eisenreich2017}. Both observations and simulations have shown that the hot halos in massive elliptical galaxies are made up of both enriched material from inside of the galaxy, as well as pristine material that flows in from outside the dark matter halo and is shock heated \citep{1992A&A...254...49A,2007ApJ...666..147G,2008SSRv..134..363S}. Naturally, the enrichment of halo gas is strongly affected by changes in the behavior of baryonic inflows and outflows. An AGN feedback model which deposits mechanical energy can effectively drive powerful nuclear winds as observed in many luminous quasars \cite[e.g.][]{2016MNRAS.459.3144Z} and can therefore have a strong effect on both the inflow and outflow of gas around a galaxy. In a recent paper,  \citet{Brennan2018} showed that the mechanical and radiation-driven AGN feedback can indeed strongly alter the baryon cycle in massive halos by comparing two sets of zoom-in simulations with and without AGN feedback. If AGN in the universe are indeed affecting the baryon cycle in their host galaxies so strongly, this is likely to leave a signature on the metal content of those galaxies, especially in the outer hot halo gas component. 

Observationally, there is accumulated evidence for metal outflows caused by AGN-driven winds \citep[see][and reference therein]{McNamara2012}. Highly metal enriched, cool gas has been recently found preferentially along the cavities or correlated with the radio jets in several clusters and groups, and found to be generally enriched to near-solar or even supersolar abundance, sometimes more enriched than regions near the central galaxies \citep{Kirkpatrick2009,Simionescu2009,Kirkpatrick2011,OSullivan2011,Kirkpatrick2015}. This metal-enriched gas is thought to be transported to its observed location by AGN-related processes due to the significant energy required to transport gas to its detected location and surprisingly tight correlations between iron radius (the maximum radius at which a significant enhancement in iron has been detected) and AGN mechanical jet power \citep{Kirkpatrick2011}.

AGN winds are also potentially important for the metal enrichment of the IGM at high redshift \citep{Khalatyan2007}, as winds propagate further from galaxies at earlier epochs when potential wells were shallower, and AGN activity was more prevalent. Ejected metals may be able to recollect into halos at later times, providing recycled fuel for galaxy growth and metal enrichment. However, few if any previous works in the literature have explicitly studied the effect of AGN feedback on the distribution and recycling of metals in the CGM and IGM.

In this paper, we study the effect of AGN driven winds on the chemical enrichment of massive galaxies and their gaseous halos. We first examine the differences in chemical enrichment of both the galaxy and the hot halo gas in two matched sets of cosmological zoom-in simulations of 27 massive halos with and without AGN feedback implemented. We also compare the simulation results to observed abundances with particular emphasis on the hot halo gas component. Finally we track the metal enrichment and transportation by AGN feedback, to study its effect on the chemical enrichment of the galaxies and beyond. We start with Section~\ref{sec:model} where we summarize the physics included in our simulations and also the tracking method we use. In Section~\ref{sec:result} we present our results, and in Section~\ref{sec:discussion} we summarize and discuss our main results. Finally, we present a summary of our main findings in section~\ref{sec:conclusions}.

\section{Simulation and Methods} \label{sec:model}
\subsection{Galaxy formation simulations}
The cosmological zoom-in simulations of galaxy formation used in this work were first presented in \citet{Choi2017}. We use two sets of high-resolution, cosmological zoom-in hydrodynamic simulations of 27 massive halos. The simulated halos have present-day total masses of $1.4 \times 10^{12}$ $ \le M_{\mathrm{vir}} /  \Msun \le 2.3 \times 10^{13}$, and present-day stellar masses of $8.2 \times 10^{10} \le M_{\ast}/ \Msun \le 1.5 \times 10^{12} $ for their central galaxies. The initial conditions are adopted from \citet{2010ApJ...725.2312O} with cosmological parameters obtained from WMAP3 \citep[][$h=0.72, \;\Omega_{\mathrm{b}}=0.044, \; \Omega_{\mathrm{dm}}=0.26, \;\Omega_{\Lambda}=0.74, \; \sigma_8=0.77 $, and $\mathrm{n_s}=0.95$]{2007ApJS..170..377S}. In this paper we use the fiducial resolution runs of  \citet{Choi2017} with the baryonic mass resolution of $m_{*,gas}=5.8 \times 10^{6} \Msun$, and the dark matter particle resolution of $m_{\mathrm{dm}} = 3.4 \times 10^{7} \Msun$. The comoving gravitational softening lengths are $\eps_{\mathrm{gas,star}} = 0.556 \rm \kpc $ for the gas and star particles and $\eps_{\mathrm{halo}} = 1.236 \rm \kpc$ for the dark 
matter. 

The simulations include star formation, supernova feedback, wind feedback from massive stars and asymptotic giant branch (AGB) stars and metal cooling and diffusion. They also include a new treatment of mechanical and radiative AGN feedback which launches high-velocity mass outflows and heats the ambient medium via X-ray radiation from the accreting black hole. We give a brief summary of the code basics and physics we include in the simulation below, but we refer the readers to \citet{Choi2017} for a detailed description of the simulations.

We note that these simulations agree with the observed stellar mass-metallicity relation \cite[see Figure 8 in][]{Choi2017} and also have already been shown to reproduce the hot-gas X-ray luminosities of galaxies in the relevant mass range, another property that has been shown to be very sensitive to the AGN feedback implementation \citep[see \citealp{LeBrun2014} and Figure 6 in][]{Choi2017}.

\subsubsection{Code basics}
The code used in this study is SPHGal \citep{2014MNRAS.443.1173H} which is a modified version of the parallel smoothed particle hydrodynamics (SPH) code GADGET-3 \citep{2005MNRAS.364.1105S}. SPHGal includes a number of improvements to overcome the numerical fluid-mixing problems of classical SPH codes have \citep[e.g.][]{Agertz2007}: a density-independent pressure-entropy SPH formulation \citep{2001MNRAS.323..743R,2013ApJ...768...44S,2013MNRAS.428.2840H}, an improved artificial viscosity implementation \citep{2010MNRAS.408..669C}, an artificial thermal conductivity \cite{2012MNRAS.422.3037R}, and a Wendland $C^4$ kernel with 200 neighboring particles \citep{2012MNRAS.425.1068D}. Finally, we also include a time-step limiter that reduces the time-step of neighboring particles around fast-moving particles to properly model shock propagation and feedback distribution \citep{2009ApJ...697L..99S,2012MNRAS.419..465D}. 

\subsubsection{Chemical enrichment, star formation and stellar feedback models}
Following \cite{2013MNRAS.434.3142A}, we include the chemical evolution and star formation model, which allows chemical enrichment by winds driven by Type~I Supernovae (SNe), Type~II SNe and asymptotic giant branch (AGB) stars. The chemical yields are adopted respectively from \citet{1999ApJS..125..439I,1995ApJS..101..181W,2010MNRAS.403.1413K} for Type I, Type II SNe and AGB stars. The masses in 11 different species, H, He, C, N, O, Ne, Mg, Si, S, Ca and Fe, are traced explicitly for star and gas particles. Then, these individual element abundances of gas particles, along with their temperature and density, are used to calculate the net cooling rate. The cooling rate is adopted from \citet{2009MNRAS.393...99W} for optically thin gas in ionization equilibrium and a redshift dependent UV/X-ray and cosmic microwave background is adopted from \citet{2001cghr.confE..64H}. We allow the metal enriched gas particles to mix their metals with neighboring gas via turbulent diffusion of gas-phase metals following \cite{2013MNRAS.434.3142A}.

We calculate star formation rate as $d \rho_{\ast} /dt = \eta \rho_{\rm gas} /t_{\rm dyn}$ where $\rho_{\ast}$, $\rho_{\rm gas}$ and $t_{\rm dyn}$ are the stellar and gas densities, and local dynamical time for gas particle respectively. We set the star formation efficiency $\eta$ to be $0.025$. Then star particles are stochastically spawned when the gas density exceeds a density threshold, $n_{\rm th} \equiv n_0 \left( T_{\rm gas}/ T_0 \right)^3 \left( M_0 / M_{\rm gas}\right)^2 $ where the critical threshold density and temperature are $n_0 = 2.0 \cm^{-3}$ and $T_0 = 12000$~K and $M_0$ is the gas particle mass in fiducial resolution. It requires that the gas density should be higher than the value for the Jeans gravitational instability of a mass $M_{\rm gas}$ at temperature  $T_{\rm gas}$.

We adopt the stellar feedback model from \citet{2017ApJ...836..204N}. The stellar feedback processes we include are the winds from young massive stars, UV heating within \strom spheres of young stars, three-phase Supernova remnant input from both type I and type II SN feedback, and outflow and metals from dying low-mass AGB stars. 

First, the young star particles distribute their wind momentum to the closest gas particles with the same amount of ejected mass and momentum as those of type II SN explosions evenly spread in time before the moment of the SN explosion. Their ionizing radiation gradually heats the neighboring gas to $T=10^4$~K within a Str\"{o}mgren radius \citep{1939ApJ....89..526S}.

When a star particle explodes as SNe, it distributes energy and momentum to the surrounding ISM, and we assume a single SN event ejects mass in an outflow with a velocity $v_{\rm out,SN}=4,500 \kms$. Depending on the physical distance from the SN particle, we assume that each neighboring gas particle is affected by one of the three successive phases of SN: (i)~momentum-conserving free expansion phase, (ii)~energy-conserving Sedov-Taylor phase where SN energy is transferred with 70\% as thermal and 30\% as kinetic, and finally (iii)~the snowplow phase where radiative cooling becomes dominant. This ``snowplow'' SN feedback model launches standard Sedov-Taylor blast-waves carrying energy as 30\% kinetic and 70\% thermal, and both energy modes dissipate with distance from the SN in pressure-driven snowplow phase of SN remnants. 

Lastly, old stellar particles keep distributing energy, momentum, and metals via AGB winds. We assume that the outflowing wind velocity of AGB stars is $v_{\rm out, AGB}=10 \kms$, which corresponds to typical outflowing velocities of AGB driven winds \citep[e.g.][]{1992A&amp;AS...93..121N}.

In summary, various mass-loss events, including winds from young stars, SNe and AGB stars, contribute to the chemical enrichment of the ISM in our model. With an assumed \citet{2001MNRAS.322..231K} IMF, over 30\% of the total mass in star particles is returned to ISM within $\sim 13$ Gyr of evolution. This ejected, metal-rich material from stars can fuel late star formation as well as AGN activity by feeding the central super massive black holes \citep[e.g.][]{2010ApJ...717..708C}. 

All total metallicities are shown in units of Solar metallicity, where we assume $Z_{\odot}$ = 0.0134 \citep{Asplund2009}.

\subsubsection{Black hole formation, growth and feedback model}\label{sec:bhmodel}
In the simulations, we seed new collisionless black hole particles with a mass of $10^5 \Msunh$ at the center of new emerging dark matter halos with mass above $1\times10^{11} \Msunh$. This choice of the dark matter halo threshold and black hole seed mass approximately follows the \citet{1998AJ....115.2285M} relation and the theoretical calculations of \cite{2017MNRAS.467.4180S}. 

We assume that the black hole can grow via two channels: mergers with other black holes and direct accretion of gas. The black hole mergers are allowed when two black hole particles are close enough within their local SPH smoothing lengths, and their relative velocities are less than the local sound speed. For the gas accretion, we adopt a Bondi-Hoyle-Lyttleton parameterization \citep{1939PCPS...34..405H,1944MNRAS.104..273B,1952MNRAS.112..195B} as we cannot fully resolve the accretion disk of the black holes on sub-pc scales in our simulation. As in \cite{2005MNRAS.361..776S}, the rate of the gas infall onto the black hole is estimated $\dot{M}_{\rm{inf}}= (4 \pi  G^{2} M_{\rm BH}^{2} \rho ) /((c_{\rm s}^2+ v^{2})^{3/2})$, where $\rho$, $c_{\rm s}$, and $v$ denote the gas density, the sound speed and the velocity of the gas relative to the black hole respectively. In addition, to prevent the unphysical infall of gas from outside of the Bondi radius, we adopt the soft Bondi criterion as in \cite{Choi2012a}, which statistically limits the accretion to the gas within the Bondi radius. It limits the accretion by the volume fraction of the gas particle lying within the Bondi radius.

Radiation from radiatively efficient accretion onto a SMBH can launch strong outflows \citep[e.g.][]{2000ApJ...543..686P,2004ApJ...616..688P}. As these wind-driving processes occur below our resolution limit, we use a sub-grid AGN-driven wind model which imparts mass and momentum to the surrounding gas following \citet{Choi2012a,Choi2014} in a manner that mimics the observed wind outflows as seen by \cite{Arav2020}.  

In this model, AGN-driven winds are launched in the vicinity of the black hole with a constant velocity $v_{\rm outf,AGN} =10,000$~$\kms$, and the wind mass is determined by a mass inflow rate and also by feedback efficiency parameter $\epsw$, which is set to 0.005 \citep{Choi2017}. The total energy flux carried by the wind is $\Edotw \equiv \epsw \Mdotacc c^2=0.5 \Mdotoutf v_{\rm outf,AGN}^2$. The mass flux and momentum flux carried by the AGN-winds are $\Mdotoutf= 2 \Mdotacc \epsw c^2 / v_{\rm outf,AGN}^2 $ and $\pdot = \Mdotoutf v_{\rm outf,AGN} = 2 \epsw \Mdotacc c^2 / v_{\rm outf,AGN}$, respectively. With our selected $\epsw$ and $v_{\rm outf,AGN}$,  90~\% of the inflowing mass is expelled as winds ($\Mdotoutf = 9 \Mdotacc$) carrying momentum flux of $\pdot = 30 L_{\rm BH} / c$. The direction of the wind is set to be parallel or anti-parallel to the angular momentum vector of each gas particle accreted by the black hole. Then, the ejected wind particle shares its momentum and energy with its two nearest neighboring gas particles to reproduce the shock heated momentum-driven flows.

In addition, we include the heating and associated radiation pressure effect from moderately hard X-ray radiation ($\sim 50$~keV) from the accreting black hole. This radiation is nearly independent of obscuration, therefore we use the AGN spectrum and Compton and photoionization heating prescription from \citet{2004MNRAS.347..144S,2005MNRAS.358..168S}. The associated radiation pressure directed away from the black hole is also included. Finally, the accretion rate onto the black holes is not artificially capped at the Eddington rate in our simulation. Instead, we include the Eddington force acting on electrons in gas particles, directed radially away from the black hole. Therefore super-Eddington gas accretion occasionally happens in our model but the corresponding feedback effects naturally reduce the inflow.

\subsection{Computing X-ray luminosities and metal abundances}
\label{sec:xraymetal}
In our simulations, the masses in 11 different species, H, He, C, N, O, Ne, Mg, Si, S, Ca and Fe, are traced explicitly for star and gas particles. We use 9 heavy elements to calculate the total metal abundance $Z$.  Among them, iron is the most commonly used metal tracer in X-ray observations due to its strong lines in soft X-ray bands. When we compare simulated gas iron abundances with observationally inferred abundances, we use the X-ray emission-weighted abundance which is calculated as  
\beq
\langle Z_{\rm Fe}\rangle = \frac{\sum j_{\nu,i} Z_{{\rm Fe},i}}{\sum j_{\nu,i} },
\eeq
where $j_{\nu,i}$ is the X-ray emissivity in the energy band of 0.3 \-- 8 keV. The X-ray emissivity is calculated following \citet{1995ApJ...451..436C} and includes Bremsstrahlung radiation and metal-line emission from all relevant species. Finally, all iron abundances are shown in units of Solar iron abundance, where we assume $\epsilon_{\rm Fe} = 7.5$ from \cite{Asplund2009}.

\subsection{Halo finders \& Particle Tracking}
We use two public codes to define and track the halo centers: the phase-space temporal halo finder, \texttt{ROCKSTAR} \citep{behroozi2013a} and \texttt{pygad} \citep{2018ascl.soft11014R}, a python module for GADGET snapshots, which adopts the shrinking sphere method to identify halos. Then we determine halo masses and halo radii using a spherical over-density threshold $\rho_{\rm th}=\rho_{200}$, 200 times that of the critical density at a given redshift, as $\Mh \equiv M_{200}$ and $\rh \equiv r_{200}$. The galaxy radius is set to 1/10 of the halo radius ($\rg=0.1 \rh$), and the star and gas masses within the galaxy radius $\rg$ are defined as the stellar and gas masses of the galaxy.

As described in \cite{Brennan2018}, we trace the flow of particles across two spherical shells of galaxy radius $\rg$ and halo radius $\rh$. The tracking begins at $z =3$ for most of our galaxies, and as early as $z \sim 4$ for a handful of galaxies with reliable centers found for both the MrAGN and NoAGN pair. We first catalogue all the gas particles that are inside of $\rg$ and $\rh$, and then check their locations and radial velocities at the next time step to determine the inflowing and outflowing masses.

\begin{figure*}
\plotone{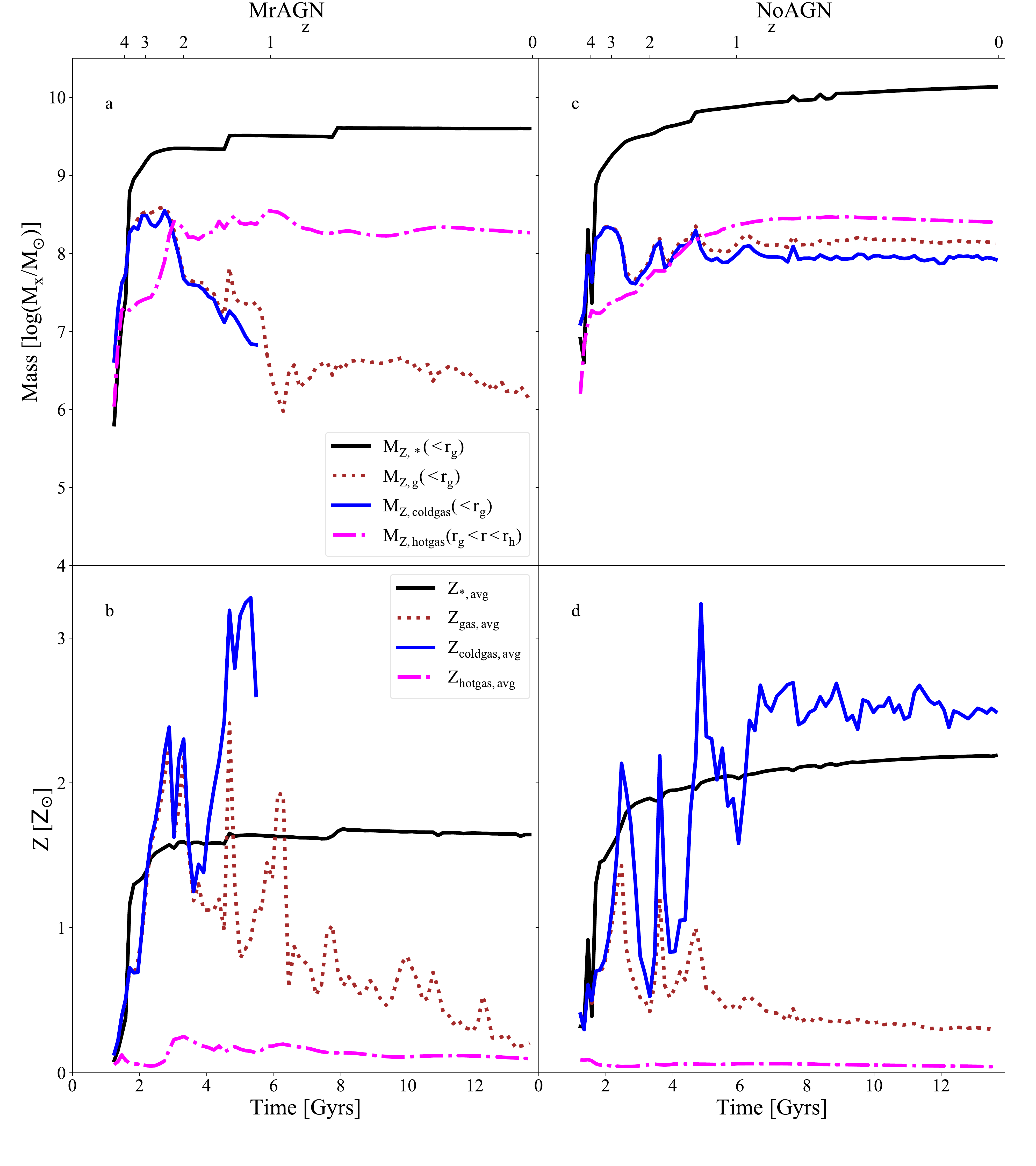} 
\caption{The time evolution of the mass in metals and the metallicity in the example galaxy m0329. We show the results of the MrAGN run in the left column (a,b), and the NoAGN run in the right column (c,d). In the upper panels, the evolution of the metal mass budget within the galaxy ($r<\rg$) in stars (black solid line), in gas (red dotted line), and in cold gas ($T < 2\times10^{4}$K, blue solid lines) are shown respectively. The evolution of mass in metals in the hot gas within the halo ($T > 2\times10^{4}$K, $\rg <r< \rh$) is shown by the pink dashed line. In the lower panels, the time evolution of the metallicity of gas, stars, cold gas, and hot halo gas in units of Solar metallicity are shown.}
{\label{metplot_329}}
\end{figure*}

\begin{figure}
\plotone{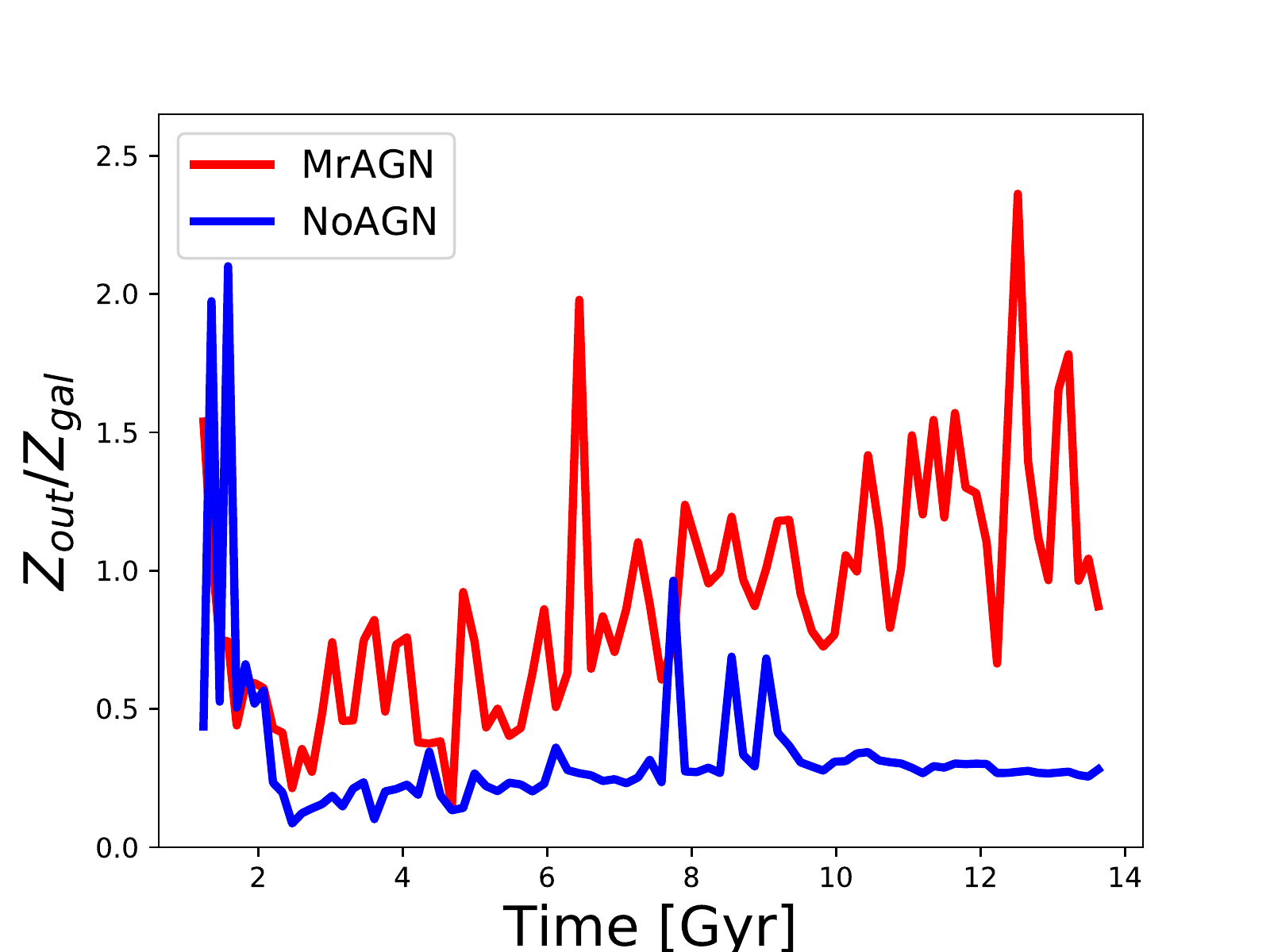}
\caption{The time evolution of the ratio between the average metallicity of gas that is outflowing across the galaxy radius $r_{g}$, i.e., 10 \% of the virial radius ($Z_{out}$) and the average metallicity of gas within the galaxy ($Z_{gal}$, $r<r_{g}$) for the example galaxy m0329.
}
{\label{out_enrich}}
\end{figure}

\section{Results} \label{sec:result}
In this section, we present results from our analysis of the simulations with and without AGN feedback. We first study trends between global properties such as the average metallicity of different galaxy components and halo mass, and compare with observed scaling relations between global quantities such as the X-ray luminosity and temperature of the hot halo gas and the iron abundance. Next we examine the impact of AGN on the spatial distribution of metals around galaxies, and compare with observations of metal abundance profiles in galaxy groups.

\subsection{Global Metallicity Trends}
\subsubsection{The history of metal enrichment: case study}
Each of our simulated halos has a different formation history. Therefore, it can be difficult to see trends when we plot time evolution for the full sample averaged together. We find it can sometimes be more illustrative to show results from a single example halo. Figure~\ref{metplot_329} shows the history of metal mass and metallicity in the representative example galaxy, m0329 ($M_{\rm{*}}\sim10^{11.3}M_{\odot}$ and $M_{\rm{h}}\sim10^{12.7}M_{\odot}$ at $z=0$).  
In the top panels we show the mass of metals contained in four different components: stars within the galaxy radius $\rg$, total gas within $\rg$, cold gas ($T < 2\times10^{4}$ K) within $\rg$, and hot gas ($T> 2\times10^{4}$ K) within the hot gas halo (CGM), between $\rg$ and the halo radius $\rh$. Since $z \sim 4$, the largest amount of metals is contained in stars within the galaxy. In the MrAGN run, AGN feedback shuts down cooling and removes most of the cold ISM from the galaxy, causing star formation to become ``quenched", while in the NoAGN run significant star formation continues all the way until $z=0$ \citep[as shown in detail for this same halo in][]{Brennan2018}. As a result, the NoAGN run produces a larger mass of metals overall by the present day. There is a slightly higher mass of metals (by $\sim0.2-0.3$ dex) in the hot halo component of the MrAGN run than in the NoAGN run, although in both runs there are more metals contained in the hot halo gas (CGM) than in the galactic (ISM) gas. 
 
In the lower panels we show the evolution of {\it metallicities} of the same four components averaged by mass. The average metallicity of stars is a bit higher in the NoAGN run, while the average metallicity of hot halo gas is larger in the MrAGN run. In both runs, the metallicity of cold ISM gas tends to be much larger than the metallicity of hot halo gas, despite the mass of metals in hot halo gas being larger: the mass of hot diffuse halo gas is much larger than the mass of cold galactic gas.

Using the particle tracking method of \citet{Brennan2018}, we can investigate the metal content of \emph{outflowing} material and how it compares to the metallicity of the ISM of the host galaxies. We track the outflowing gas mass across the galaxy radius $r_{g}$, i.e., 10 \% of the virial radius of the galaxy. Then we check the metal abundance of the outflowing mass and its ratio to the metal abundance of the gas within the galaxy ($r < r_{g}$). 

In Figure~\ref{out_enrich} we show the evolution of the ratio of the average metallicity of outflowing gas to the average metallicity of gas within the example halo m0329 with time. The outflowing gas leaving the galaxy across the galaxy radius $r_{g}$ in the MrAGN run is more enriched relative to the rest of the gas in the galaxy than the outflowing gas in the NoAGN run. The metallicity of outflowing gas at certain times can be even higher than the average metallicity of all of the gas left behind ($Z_{\rm out} / Z_{\rm gal} > 1$). In both runs, this ratio increases with time, but the increase is more pronounced in the MrAGN run. 

\subsubsection{Mass and Metal Budget: Broad Trends}
\begin{figure*}
\plotone{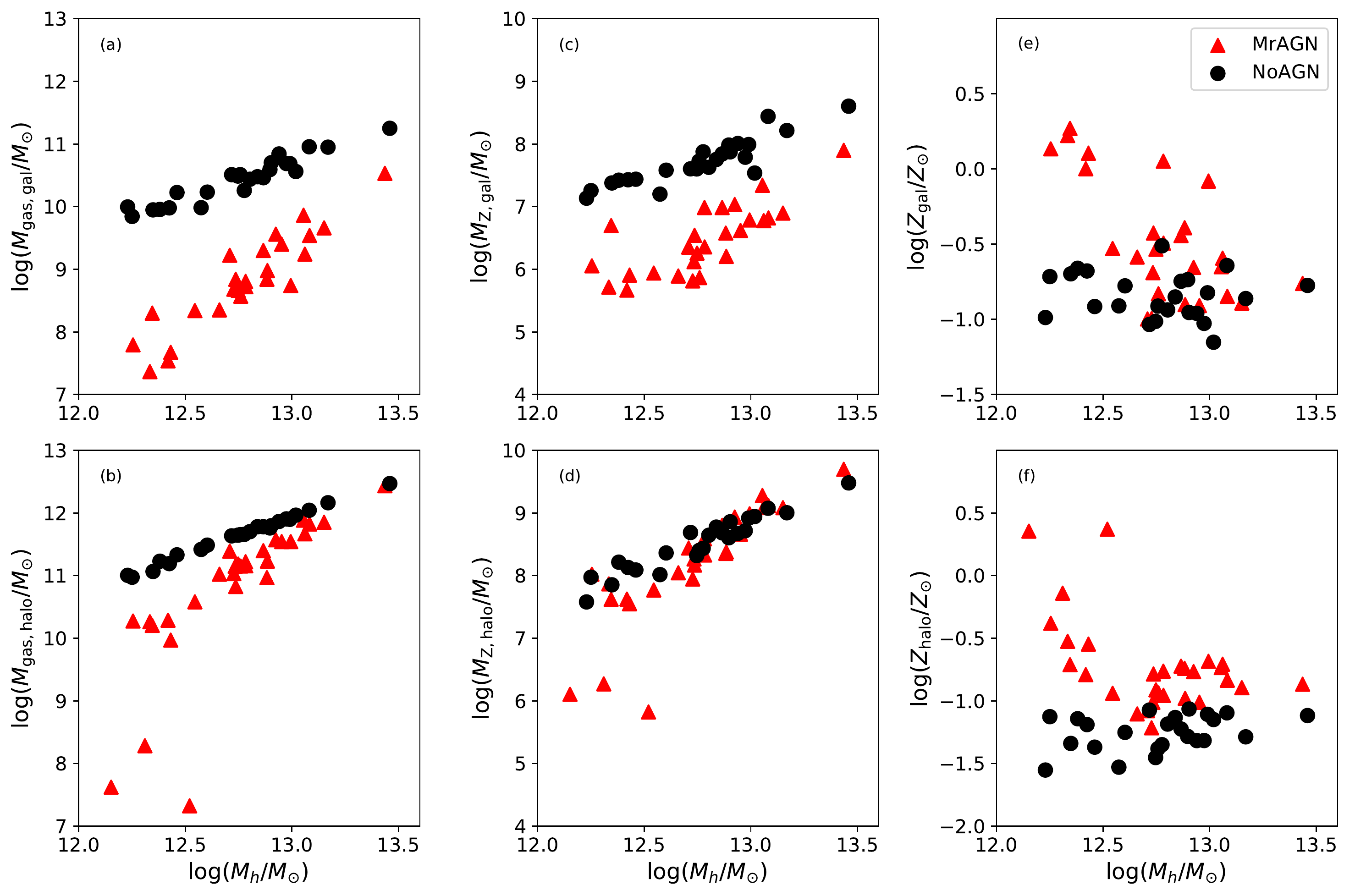}
\caption{The total mass, total metal mass, and the metallicity of hot gas ($T>2\times10^{4}$ K) within the galaxy radius, $\rg$ (top panels) and within the halo, $\rh$ (bottom panels) are shown as a function of halo mass at $z=0$. The galaxies simulated with AGN feedback (MrAGN) are shown by red triangles and the galaxies simulated without AGN feedback (NoAGN) are shown by black circles respectively. Overall, MrAGN galaxies have less hot halo gas compared to NoAGN galaxies, and the differences are larger on the galaxy scale.}
{\label{gasbudget}}
\end{figure*}

We now turn to our full suite of simulated halos. Figures~\ref{gasbudget} summarizes the gas and metal budget in the hot phase ($T>2\times10^{4}$K)  in our simulated galaxies within the galaxy ($r < \rg$) and within the halo ($r < \rh$) at $z=0$. In the left panels, we first show the masses of hot gas  contained in both the galaxy and the halo of all of our simulations at $z=0$. In the panel (a) of Figure~\ref{gasbudget}, we see that the overall gas mass inside galaxies is greatly reduced in the MrAGN run when compared with the NoAGN run, especially at low mass where the difference can be as great as 2 dex. In panel (b), we show that the mass of gas in the halo is also decreased by as much as 1 dex in lower mass galaxies. AGN feedback removes gas from both galaxies and halos in all galaxy mass ranges but with more pronounced differences between the two runs in lower mass halos \citep[see also][]{Brennan2018}. 

Turning to the middle panel (Figure \ref{gasbudget}-(c) and (d)), the mass of metals in MrAGN galaxies is reduced, but by a slightly smaller percentage than the total gas mass as shown in the left panels (a) and (b). Meanwhile, panel (d) shows that the mass of metals in the halo is very similar between the two runs, despite the smaller mass of gas in the halo found in the MrAGN run. Therefore as shown in the right panels, (e) and (f), this manifests as slightly \emph{higher} gas metallicities in MrAGN galaxies both within the galaxies and in the halos. This trend is slightly more pronounced in lower mass halos.

\begin{figure}
  \plotone{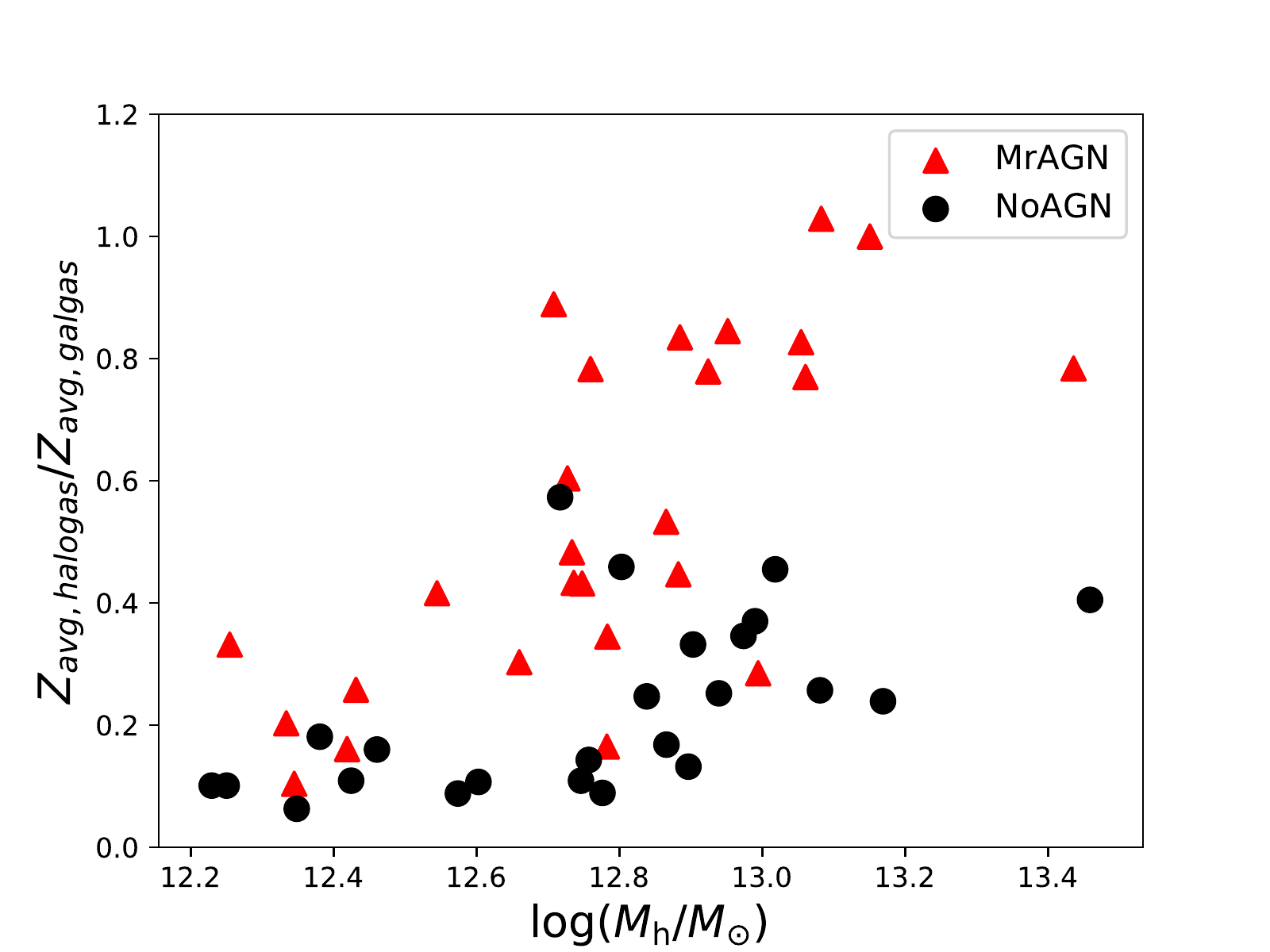}
  \caption{Ratio of average metallicity of halo gas to average metallicity of 
  galactic gas within the galaxies at $z=0$. The galaxies simulated with 
  AGN (MrAGN) are shown by red triangles and the galaxies simulated 
  without AGN feedback (NoAGN) are shown by black circles.}
          {\label{metallicityratio}}
\end{figure}

Figure \ref{metallicityratio} shows the ratio of average halo gas metallicity to the average metallicity of gas within the galaxy for our full sample at $z=0$. In all cases, this ratio is higher for the MrAGN runs, with the difference between runs being greatest for higher mass galaxies, where the ratio approaches and even slightly exceeds unity. We can think of this as the cumulative effect of the metal enriched outflows seen in Figure~\ref{out_enrich}. Ejected metals tend to remain within the halo in massive systems due to their deeper potential wells, cumulatively enriching the halo gas, while most of the ejected gas escapes beyond the virial radius of the smaller halos.

\subsubsection{Iron abundance scaling relations compared with observations} 
\begin{figure}
\plotone{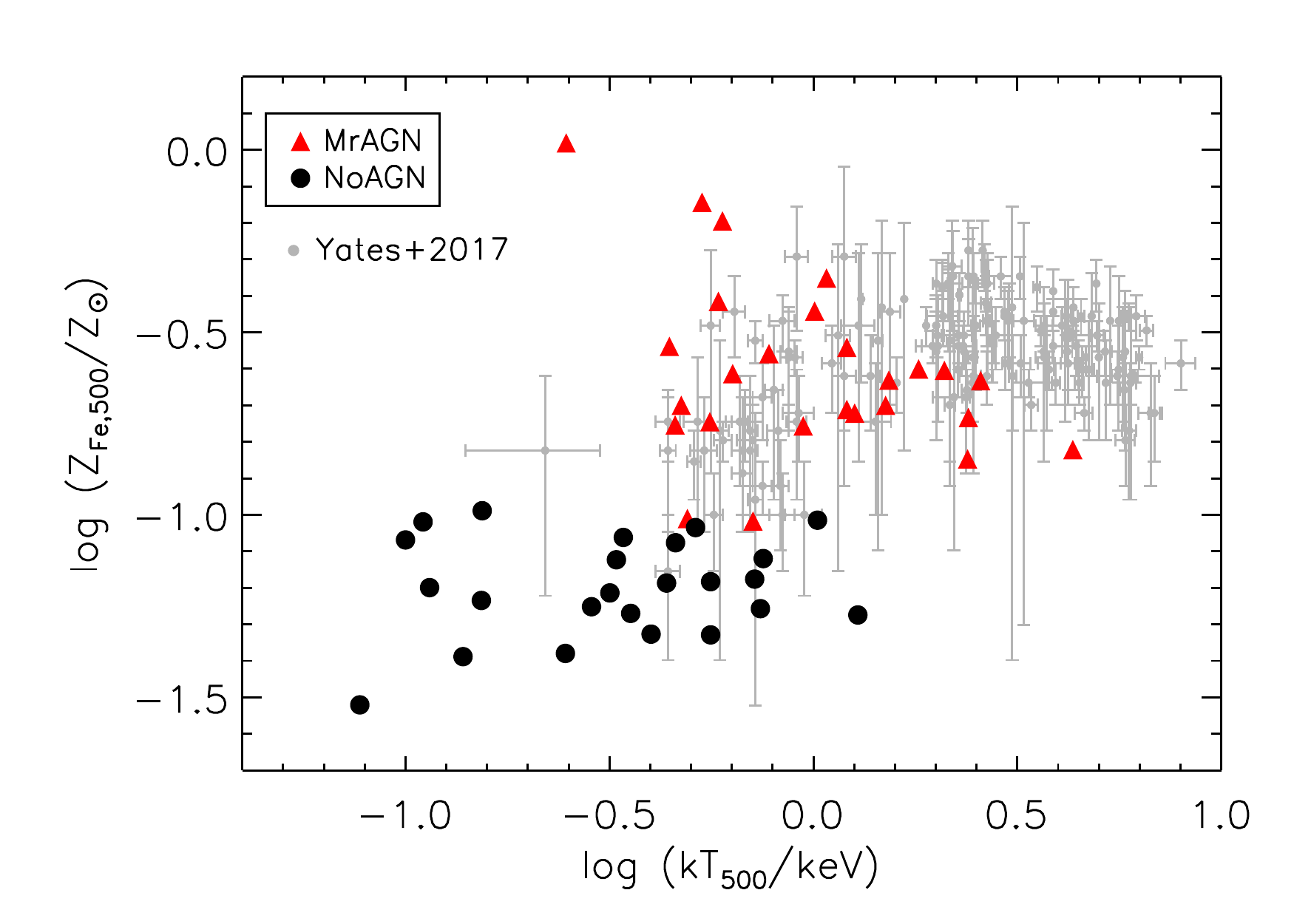}
  \caption{The iron abundance of hot halo gas versus X-ray temperature of 
  our simulated galaxies with and without AGN feedback. The galaxies simulated with AGN 
  (MrAGN) are shown by red triangles and the galaxies simulated without AGN 
  feedback (NoAGN) are shown by black circles. Observational estimates are shown by grey circles (see text for details).
  The MrAGN simulations produce X-ray properties of group-sized halos that are in better agreement with observations than the NoAGN simulations. }
{\label{fe_lumkt}}
\end{figure}

Figure \ref{fe_lumkt} shows the \emph{mass-weighted} iron abundance of gas within $\rm{R_{500}}$ versus the X-ray-emitting-gas temperature within $\rm{R_{500}}$ of the simulated galaxies. We compare the two models with a homogenized dataset of low-redshift groups and clusters compiled by \citet{2017MNRAS.464.3169Y}. In this analysis, the authors have converted the observed X-ray emission-weighted iron abundance into a mass-weighted iron abundance within a standardized aperture, using an assumed template temperature and iron abundance profile. The observed k$T_{500}\--Z_{\rm Fe,500}$ relation from  \citet{2017MNRAS.464.3169Y} is fairly flat, with increased scatter in $Z_{\rm Fe,500}$ below log (k$T$/keV) $\sim 0.2$. There is a slight negative trend in k$T_{500}\--Z_{\rm Fe,500}$ for systems with log (k$T$/keV) $\gtrsim 0.3$.
The MrAGN simulations show much higher iron abundances than the NoAGN run and roughly reproduces the iron abundances measured for the clusters with log (k$T$/keV) $< 0.2$. However, the halos with the highest temperatures in our MrAGN suite show iron abundance on or slightly below the lower envelope of the observed sample. On the other hand, galaxies simulated without AGN feedback have much cooler temperatures and lower iron abundances than the observed sample for their halo mass range.

\begin{figure*}
\plotone{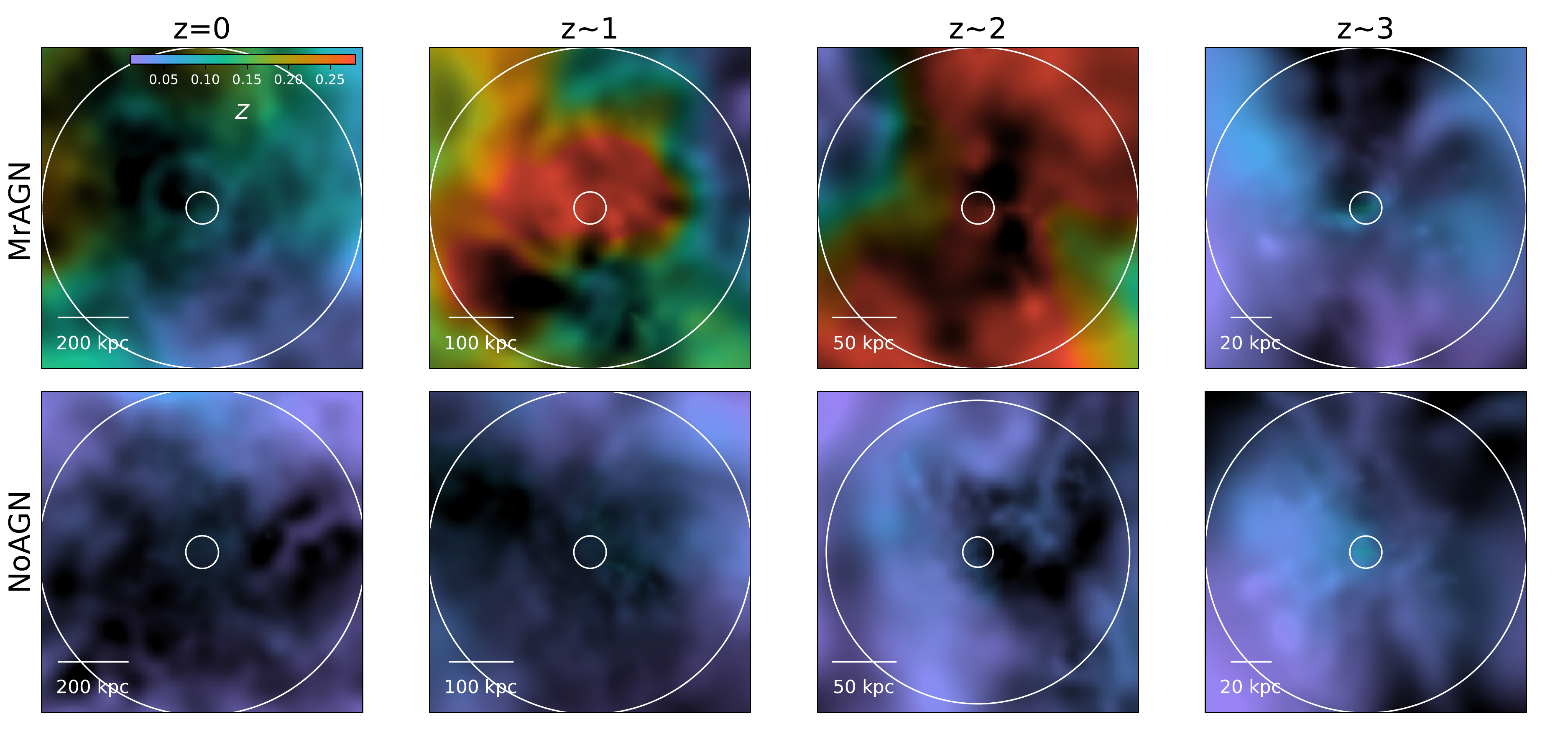} 
\caption{Gas metallicity maps at $z=0,1,2$ and 3 for the example galaxy m0329. The color corresponds to gas metallicity in Solar units, and the luminance of the image corresponds to gas density. The inner/outer white circles denote the  galaxy radius (10 \% of the virial radius) $\rg$ and the virial radius $\rh$ at each redshift respectively. The MrAGN case is shown in the top panels and the NoAGN case in the bottom panels. There is evidence of a large-scale bipolar chemically enriched outflow depositing metals into the halo at $z=2$ in MrAGN case. The final halo gas metallicity at $z=0$ is overall higher in MrAGN case than in NoAGN. The images are created with \texttt{pygad} \citep{2018ascl.soft11014R}. \label{vel_vec_met_329}}
\end{figure*}

\subsection{Where are the Metals: the spatial distribution of metals around galaxies}

\begin{figure*}
\plotone{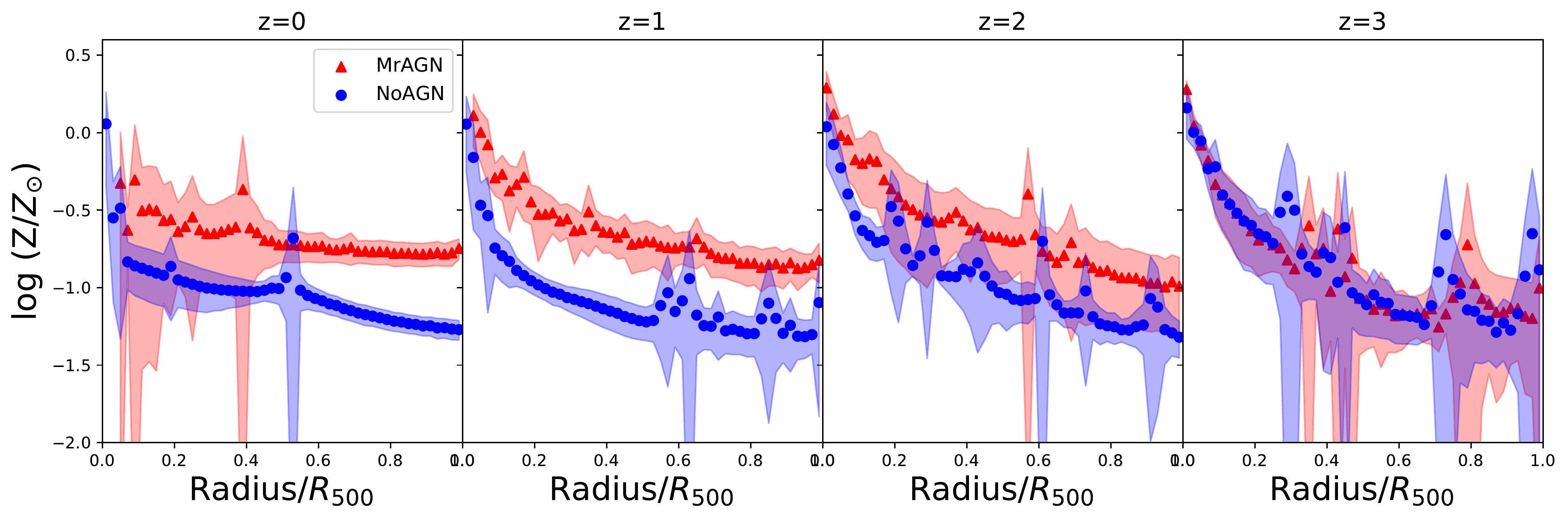}
\caption{Average gas metal abundance profile for all gas in halos with $M_{\rm vir} > 8\times10^{12}M_{\odot}$ at redshift $z=0,1,2$ and $3$. The MrAGN  and NoAGN cases are shown by red and blue symbols respectively, and the shaded regions represent the standard deviation over the suite of different halos.  Galaxies simulated with AGN winds start to show overall higher gas metallicity out to $R_{500}$ since $z=2$, and end up with flatter abundance profiles than the simulations without AGN feedback.
\label{fe_grad_stack}}
\end{figure*}

\subsubsection{Metal Maps: Case Study}
Once again, the detailed distribution of metals around galaxies varies greatly from one halo to another, so we again use the case study halo m0329, simulated with and without AGN feedback, to illustrate how AGN feedback affects the metal distribution in the CGM. In Figure~\ref{vel_vec_met_329}, we present the metallicity map of the gas in halo m0329 at four different redshifts $z=0,1,2$ and $3$. The metallicity is again shown in units of solar metallicity. At $z=3$, the distribution of metallicity looks very similar between the two runs with and without AGN feedback, but after $z=2$ the outflow signatures can be seen in the MrAGN case. In the snapshot at $z=2$, there is evidence of a large-scale, biconical, chemically enriched outflow depositing metals into the galactic halo in the MrAGN run. These metal-enriched bipolar winds are driven from the central supermassive black hole. Large scale cavities also form along the direction of bipolar AGN-driven winds, as winds efficiently blow out existing halo gas as they propagate. It is also notable that the \emph{inhomogeneity} of the metal distribution in the MrAGN case is much greater. This might be possible to test with observations.

\begin{figure}
\plotone{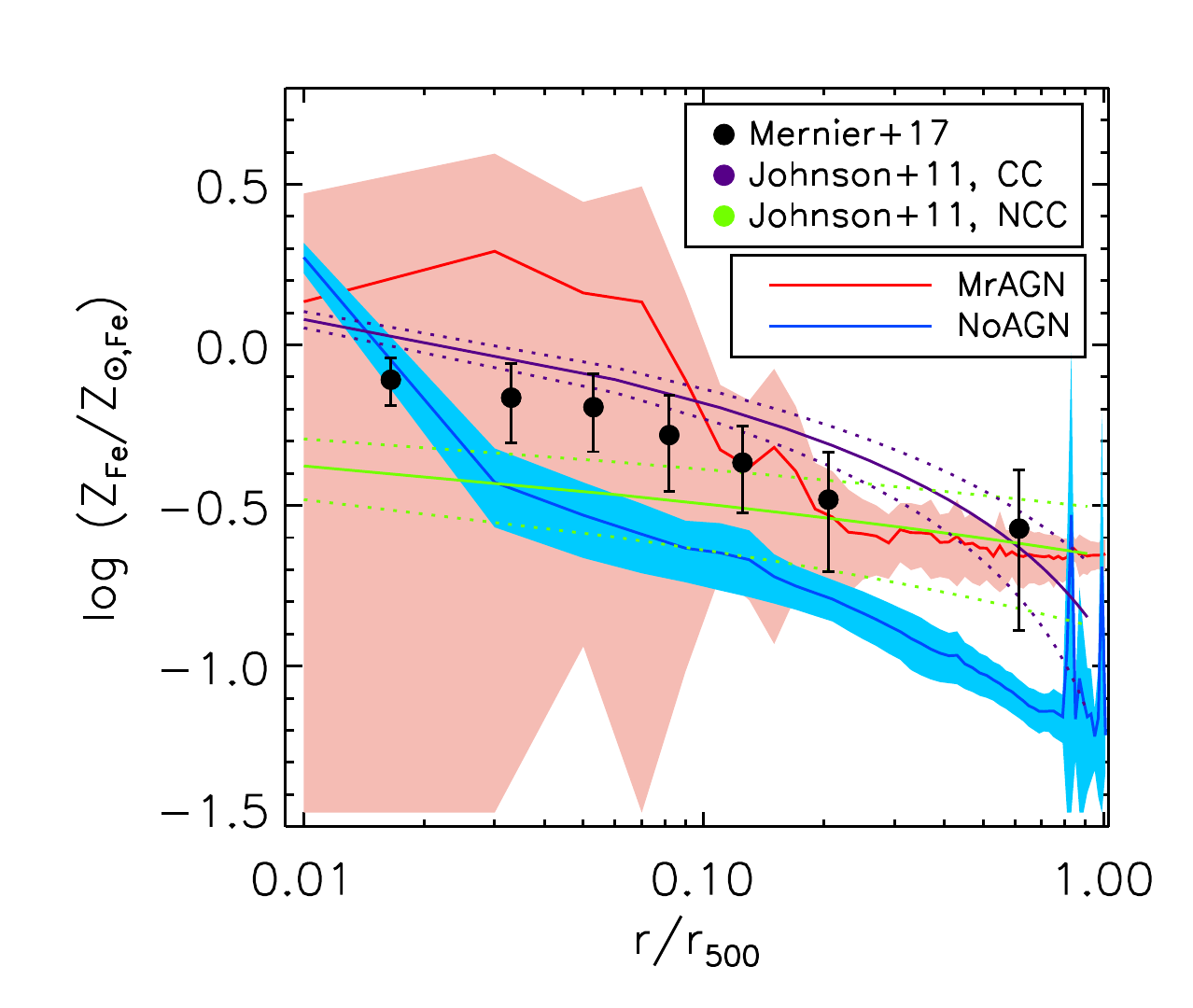}
  \caption{X-ray emission weighted iron abundance profiles of simulated galaxies with $M_{\rm vir}> 8\times10^{12}M_{\odot}$. MrAGN  and NoAGN cases are shown by red and blue lines respectively, and the color shaded regions represent the standard deviation over the suite of halos. Green and purple solid and dashed lines show the stacked profiles from the study of \citet{2011MNRAS.413.2467J} for non-cool core and cool core groups, respectively. Black symbols show the averaged radial iron abundance profile of 21 nearby cool core groups \citep{Mernier2017}. } 
  {\label{fe_profile}}
\end{figure}

\subsubsection{Stacked Metallicity Abundance Profiles}
We first present the stacked (mass weighted) metal abundance profile for all of our massive halos with $M_{\rm halo} >8\times10^{12} \Msun$ in Figure~\ref{fe_grad_stack}. It shows the radially averaged metal abundance profile for {\it all gas} at four different redshifts for the MrAGN and NoAGN runs. The two models started with very similar abundance profiles at $z=3$, but diverge by $z=2$ as frequent AGN activity starts to drive out enriched material. By $z=0$, enriched material from the central part of the galaxy has been redistributed outwards, resulting in flatter gas abundance profiles and higher gas metal abundances in the outer parts of the halo $r>0.5 {R_{500}}$. The overall gas metal abundance across the halo at $z=0$ ends up being higher in MrAGN run. There is a large halo-to-halo scatter in the metal abundance in the central regions, particularly in the MrAGN runs. The large spikes in the profiles are due to the metal-rich gas associated with satellite galaxies.

We now turn to a comparison of the predicted iron abundance profiles to recent observations. Figure~\ref{fe_profile} shows X-ray emission weighted and projected iron abundance profiles of simulated galaxies with $\rm{M_{\rm vir}}$ $> 8 \times 10^{12}M_{\odot}$, which is comparable to the mass range in observed samples. As shown previously in Figure~\ref{fe_grad_stack}, the average iron abundance profile of the NoAGN runs shows overall lower iron abundance at $r > 0.02 R_{500}$, but higher abundance for the innermost region.  On the other hand, MrAGN galaxies show overall higher and flatter average abundance profiles at large distance ($r > 0.1 r_{500}$) in good qualitative agreement with observations. At $0.5 r_{500}$, the predicted mean iron abundance of the MrAGN galaxies is $Z_{\rm Fe} = 0.233$ $Z_{\rm Fe,\odot}$, while the mean abundance of NoAGN galaxies is 0.093.

\subsubsection{The fate of ejected metals}

\begin{figure}
  \plotone{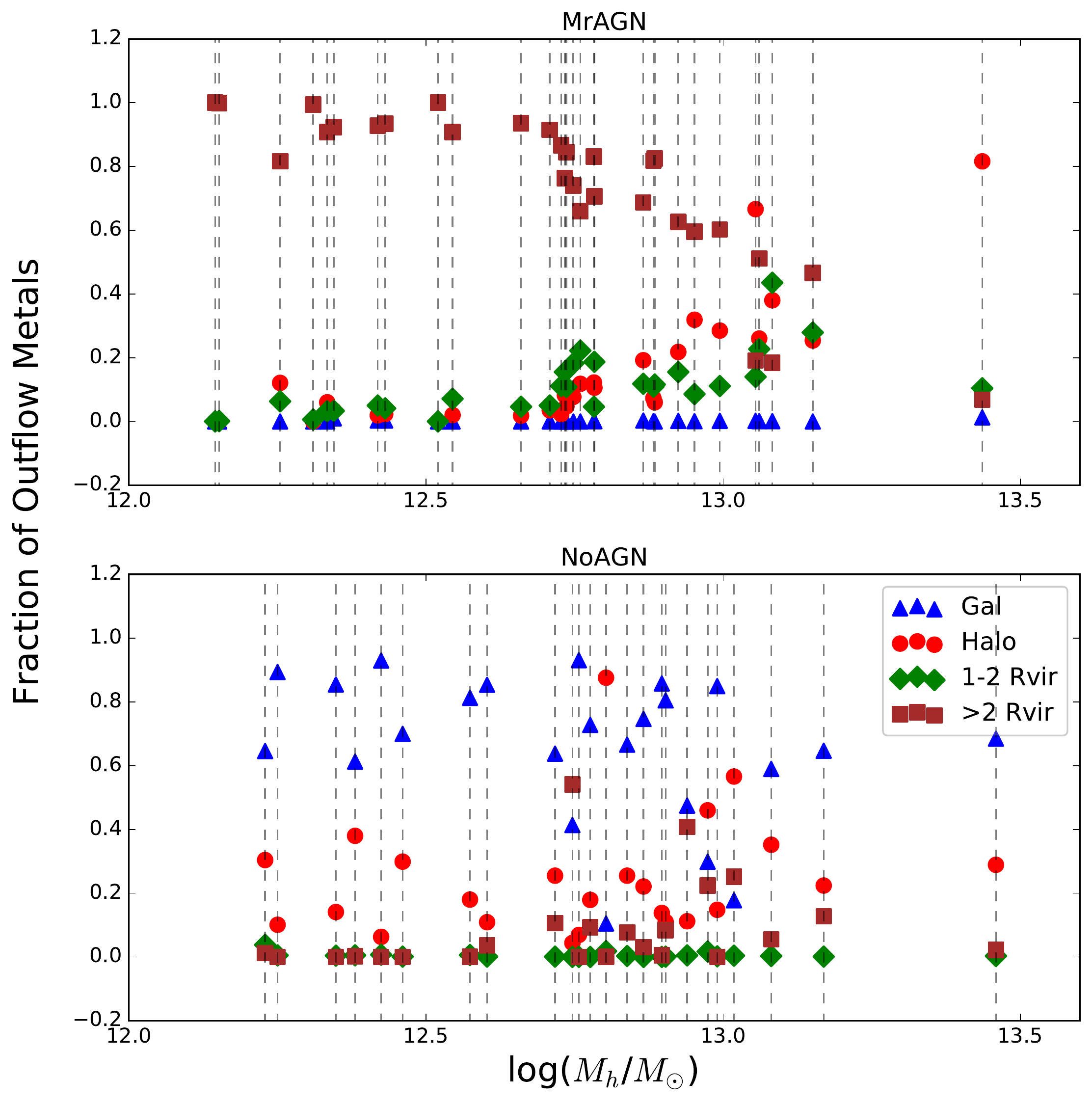}
\caption{The fraction of metals which have been ejected from the galaxy that occupy a set of radial bins at $z=0$. Each black dotted line represents a galaxy. Blue triangles denote the fraction of ejected metals which are now back in the galaxy. Red circles denote ejected metals now in the halo, within the virial radius. Green diamonds denote ejected metals that now reside between one and two virial radii, and brown squares denote metals at distances greater than two virial radii.}
{\label{outflow_budget}}
\end{figure}

We have seen that AGN can re-distribute metals from the central parts of galaxies out into the circumgalactic hot halo. However, \cite{Brennan2018} showed that much of the ejected gas driven from the galaxy ends up \emph{beyond the halo virial radius}. We therefore now explore the fate of ejected metals and where they reside at $z=0$. Figure \ref{outflow_budget} summarizes the final destination of the ejected metals, showing the percentage of all of the metals which have been {\it ejected} from each galaxy in a set of radial bins at $z=0$; galaxy ($r < \rg$), halo ($r < \rh$), within the turn-around radius $\rh < r < 2 \rh$, and finally outside the turn-around radius ($r >  2 \rh$). In MrAGN galaxies, the vast majority of metals driven from the galaxy do not return to the halo and reside as far as or farther than 2 virial radii from the galaxy center. Towards higher masses, a larger fraction of metals ejected from the galaxies come back, and exist within the halo. In NoAGN galaxies, a much larger fraction of metals returns to the galaxy and ends up in the ISM and stars, with most of the rest occupying the halo in the hot gas phase.

This shows that AGN winds are indeed potentially important for the metal enrichment of the IGM. The {\it fraction} of ejected metals that has escaped the halo declines with increasing halo mass, but still, total metal mass output from these massive galaxies can be very significant for polluting the IGM. Due to the limited volume of the zoom-in simulations presented in this paper, we reserve the quantitative study of IGM enrichment via AGN-driven winds to a future work.

\section{Discussion}\label{sec:discussion}


\subsection{The impact of AGN-driven winds on the mass and metal content of galaxies and halos}
At high redshift ($z\gtrsim3$), before black holes have grown significantly, the mass assembly history, chemical history, and baryon cycle are very similar in the MrAGN and NoAGN runs \citep[see also][]{Brennan2018}. When the central black hole becomes large enough, it starts to drive energetic winds which remove both gas and metals from the nucleus of the galaxy and disperse this material to larger radii. The mass of metals in galactic gas, $M_{Z, \rm g} (<\rg)$, is lower by more than 1 dex in the MrAGN run when compared with the NoAGN run, due to two main reasons. First, as already noted, fewer stars are born to enrich their surrounding gas with AGN feedback. Secondly, both gas and metals are removed from the ISM by AGN-driven outflows. The outflows from the galaxy can be quite enriched compared with the average value in the ISM, as winds preferentially remove material from the central parts of the galaxy, which tends to be more enriched than that in the outskirts. Thus, several of the lower mass galaxies have significantly higher gas phase metallicities in the MrAGN runs than in the NoAGN runs, although for the massive halos in many cases the galaxy scale gas metallicities are nearly the same in the two runs. This highly metal enriched outflowing gas found in the MrAGN run is consistent with some recent observations showing evidence of  metal-enriched outflows compared to the ISM of host galaxies \citep{OSullivan2011,Chisholm2016}. Recently, \cite{Gan2019} also found that broad absorption line wind regions show much higher metal abundance, up to $\sim 8 Z_{\odot}$.

The winds also entrain large amounts of hot gas in the halo, leading to a decrease in both the mass and metal mass in the hot halo relative to the total halo mass (Fig.~\ref{gasbudget}). This not only pushes gas out of the halo, but also \emph{reduces} pristine inflows of gas into the halo \citep{Brennan2018}. The prevention of inflowing metal-poor gas in the MrAGN runs is likely the main reason that the overall metallicity of the hot gas ends up being a bit higher in the MrAGN runs. 
The ejection of metals from the galaxy into the halo also leads to a higher ratio of halo gas metallicity to galaxy metallicity in massive halos (Figures~\ref{metallicityratio}). In the less massive halos, the majority of ejected metals end up far outside of the virial radius of the halo (Figure \ref{outflow_budget}).

In addition to raising the overall metal abundance of the hot halo gas, AGN-driven winds transport metals outwards from the centers of the halo, leading to flatter abundance profiles in the intragroup medium, in good qualitative agreement with observations of non-cool core groups (Figure \ref{fe_grad_stack}). The simulations with AGN feedback also produce much more inhomogeneous metallicity maps in the CGM, which may be possible to test with observations in the future. 

\subsection{ Comparison with Literature Studies}

Using the \textsc{gimic} simulations, \cite{Crain2013} studied the iron abundance profile of $L_{\ast}$ galaxies and  found a negative radial iron abundance gradient in the absence of AGN feedback. Their predicted iron abundance is around $Z_{\rm Fe} \sim 0.1 Z_{\rm Fe,\odot}$ at $0.5 r_{200}$, which is comparable to the mean abundance of our NoAGN galaxies and lower than estimates from X-ray observations. 

Simulations including AGN feedback tend to find better agreement of iron abundance gradient to observations. \cite{Fabjan2010} studied the effect of AGN feedback on the metal enrichment of galaxy clusters and found that the simulations of galaxy cluster and group formation with AGN feedback predict a flatter iron abundance profile than the ones without AGN feedback, especially at larger radii. 

\cite{Pellegrini2019} recently also found good agreement of the iron abundance profile of the simulated gas to observations in the outer region of the galaxies using the MACER simulations, high resolution idealized simulations of a massive elliptical galaxy.  The simulation includes both  SN and AGN feedback, however, they still overpredict the iron abundance in the innermost region.

\cite{Barnes2017} studied the metallicity distribution of the reference model of the \textsc{eagle} simulation as well as 30 cosmological zoom simulations of galaxy clusters from the Cluster-EAGLE (\textsc{c-eagle}) project. The mass scale of \textsc{c-eagle} is  $M_{200} = 10^{14\--15.4} M_{\odot}$, which is higher than our simulations. They found the total metal content of the simulated clusters and groups in the \textsc{eagle} simulations are a good match to the observed metallicities from \cite{2017MNRAS.464.3169Y}. They showed that the \textsc{eagle} and \textsc{c-eagle} simulations predict a relatively flat metallicity-temperature relation, which is consistent with our result shown in Figure~\ref{fe_lumkt}. They also examined the iron abundance profiles of the \textsc{c-eagle} clusters and find overall reasonable agreement between the observed profile and the median profile of the \textsc{c-eagle} sample, which is simulated with a modified stochastic AGN feedback model from \cite{Booth2009}. The simulated iron abundance profile is marginally steeper than observed, but the difference is within the intrinsic scatter.

\subsection{Caveats and Limitations of Simulations}
Overall the predicted metal enrichment of the simulation with mechanical and radiative AGN feedback model presented in this paper shows a fairly good agreement to the observed data, but some small discrepancies still remain. The excess of hot gas phase metals in the inner region of halos may be due to physical mechanisms that have not been included in the simulation, such as cosmic-ray driven winds, and relativistic AGN jets. Also, due to the limited volume of the zoom-in simulations we limited our analysis to within 2 virial radii from the galaxy center. 


\section{Conclusions}\label{sec:conclusions}
We investigated the effect of mechanical and radiation-driven AGN feedback on the metal content of massive galaxies and their gaseous halos by following the history of chemical enrichment and outflows in a suite of high resolution  zoom-in simulations. Our main findings are summarized as follows.
\begin{itemize}
    \item The prescription for AGN feedback via radiation driven winds presented by \citet{Choi2017} is capable of ejecting large amounts of metals from the galaxy into the halo and beyond.
    \item Simulations without AGN feedback produce more metals overall, but the radial metallicity profile of hot gas is too centrally concentrated and too low in the outskirts compared with observations. The simulations with AGN feedback are in good agreement with observed metallicity profiles. 
    \item Simulations with AGN feedback produce better agreement with observed scaling relations between the iron abundance of hot halo gas and X-ray luminosity and temperature than NoAGN simulations, although there is still some tension with observations. 
    \item AGN driven winds result in a slight decrease in the metallicity of gas within galaxies, and a notable increase in the metallicity of their halo gas.
    \item In simulations with stellar feedback only (NoAGN), 60-80\% of ejected metals end up back in the galaxy by the present day, and less than 10\% of ejected metals are found outside the virial radius at $z=0$. In contrast, in simulations with AGN-driven winds, only a few percent of ejected metals end up within the galaxy, and 80-90\% are outside the virial radius. The fraction of ejected metals that remains outside the halo virial radius at $z=0$ is a strong function of halo mass, decreasing from $\gtrsim 90$\% in our lowest mass halos to $<10$\% in our highest mass halo. 
\end{itemize}

\acknowledgments
We are grateful to the anonymous referee for helpful comments. We thank Silvia Pellegrini and J. Xavier Prochaska for helpful conversations. Numerical simulations were run on the computer clusters of the Princeton Institute of Computational Science and engineering. The authors acknowledge the Office of Advanced Research Computing (OARC) at Rutgers, The State University of New Jersey for providing access to the Amarel cluster and associated research computing resources that have contributed to the results reported here. URL: http://oarc.rutgers.edu. RSS acknowledges support from the Simons Foundation. MH acknowledges financial support from the Carlsberg Foundation via a ``Semper Ardens'' grant (CF15-0384).

\bibliography{library}
\end{document}